\documentclass{article}
\usepackage{fullpage}
\usepackage{cite}
\usepackage{amsmath,amssymb,amsfonts}
\usepackage{graphicx}
\usepackage{fancybox}
\usepackage{textcomp}
\usepackage{xcolor}
\usepackage{url}
\usepackage{tabularx}
\usepackage[colorlinks]{hyperref}
\hypersetup{
     colorlinks   = true,
     linkcolor    = blue,
     urlcolor     = blue,
     citecolor    = blue
}
\usepackage{comment}
\newcommand{\todo}[1]{\textcolor{red}{#1}}

\renewcommand{\todo}[1]{}
\usepackage{flushend}
\begin{document}

\title{Implementation-Oblivious Transparent Checkpoint-Restart for MPI}

\author{
Yao Xu \\
Khoury College of Computer Sciences \\
Northeastern University \\
Boston, MA, USA \\
xu.yao1@northeastern.edu
\and
Leonid Belyaev \\
Khoury College of Computer Sciences \\
Northeastern University \\
Boston, MA, USA \\
belyaev.l@northeastern.edu
\and
Twinkle Jain \\
Khoury College of Computer Sciences \\
Northeastern University \\
Boston, MA, USA \\
jain.t@northeastern.edu
\and
Derek Schafer \\
\hspace*{1.1cm}University of New Mexico\hspace*{1.1cm} \\
Albuquerque, NM, USA \\
dschafer1@unm.edu
\and
Anthony Skjellum \\
\hspace*{1.4cm}College of Engineering\hspace*{1.4cm} \\
Tennessee Tech University \\
Cookville, TN, USA \\
askjellum@tntech.edu
\and
Gene Cooperman \\
Khoury College of Computer Sciences \\
Northeastern University \\
Boston, MA, USA \\
gene@ccs.neu.edu
}

\date{}

\maketitle

\begin{abstract}
This work presents experience with traditional use cases of
checkpointing on a novel platform.  A single codebase (MANA) transparently
checkpoints production workloads for major available MPI implementations:
``develop once, run everywhere.''  The new platform enables application
developers to compile their application against any of the available
standards-compliant MPI implementations, and test each MPI implementation
according to performance or other features.

Since its original academic prototype, MANA has been under development
for three of the past four years, and is planned to enter full
production at NERSC in early Fall of 2023.  To the best of the
authors' knowledge, MANA is currently the only production-capable,
system-level checkpointing package running on a large
supercomputer (Perlmutter at NERSC) using a major MPI implementation
(HPE Cray~MPI).  Experiments are presented on
large production workloads, demonstrating low runtime overhead with
one codebase supporting four MPI implementations: HPE~Cray~MPI, MPICH,
Open~MPI, and ExaMPI.
\end{abstract}

\section{Introduction}

A checkpointing library for MPI-agnostic checkpointing of MPI applications has been available for application-level checkpointing since at least 2003~\cite{bronevetsky2003automated,schulz2004implementation}.  This allows an MPI developer to choose the best MPI implementation for that application.  Similarly, some large application codes maintain their own code for saving and restoring data across a checkpoint.   
As with all application-level approaches, the developer has the burden of declaring and maintaining those data structures that must be preserved and restored across a checkpoint-restart.

However, many large, complex codes do not have their own application-level checkpointing.  One large application that is typical of this constraint is VASP~\cite{hafner2008ab}.
VASP accounted for approximately 20\% of CPU time at the NERSC supercomputing center~\cite{NERSC} as of 2020~\cite[Figure~4]{driscoll2020automation}.
VASP supports multiple algorithms and data structures that are continually evolving.
As the VASP code evolves, it would be a large burden to continually update any application-specific module to reflect VASP's latest algorithms and data structures.

MANA~\cite{garg2019mana} (MPI-Agnostic Network-Agnostic
checkpointing) is open-source, and freely available at
\url{https://github.com/mpickpt/mana}.  MANA~\cite{garg2019mana}
has achieved production quality in recent testing, and is planned for
production use at NERSC~\cite{NERSC} in early Fall of 2023.  To the best
of the authors' knowledge, MANA is currently the only production-capable,
\emph{system-level}
checkpointing package for MPI~\cite{xu2021mana}.

Yet, just as it would be a burden for each application developer to implement their own application-specific checkpointing for MPI, it is also a burden for each MPI implementation to support its own transparent checkpointing feature.  This was attempted in the decade of the 2010s for
MVAPICH and Open~MPI (see details in Section~\ref{sec:relatedWork}). However, those modules required specific code to support each type of network being used by MPI.  Eventually, the need to support checkpointing of a proliferation of networks~\cite{lu2022survey,mellanox-infiniband,birrittella2015intel,cray-aries,desensi2020depth} caused the MPI developers to drop checkpoint support.

MANA's split-process approach (Section~\ref{sec:background:splitProcess}) created, for the first time, a transparent checkpointing package that was Network-Agnostic in the strong sense of not requiring any knowledge or code to support particular networks.  Earlier MPI-specific checkpointing implementations relied on knowledge of each supported network, in order to shut down MPI's network connections prior to a checkpoint and then restart the network connections during restart.  Hence, although Open~MPI promised ``Interconnect-Agnostic'' checkpointing~\cite{hursey2009interconnect}, it still required sufficient network code for stopping and starting each supported network.

In addition to transparent and application-level checkpointing, one should note the intermediate choice of library-based, application-specific checkpointing.  For examples, see VeloC, SCR, FTI, ULFM, and Reinit in Section~\ref{sec:relatedWork}.  These libraries are often used in conjunction with applications that support a particular execution model:  a main loop, with each iteration globally synchronized by a call to an MPI barrier.  At the synchronization point, there are no active MPI objects, and hence no MPI context needs to be saved.

The library-based and transparent varieties of checkpointing each make important contributions.  First, transparent checkpointing will usually be preferred for applications that do not follow the execution model of a globally synchronized main loop.  As stated, VASP accounts for about 20\% of execution time at NERSC, and yet its multi-algorithm execution model conflicts with the model of a single main-loop often assumed by library-based packages.

Second, an MPI application may encounter bugs and crash at an arbitrary place in the code.  The ability to take frequent checkpoints at arbitrary times aids in isolating the environment for analysis and replay, just prior to the crash.

Third, and most important for the future, the United States DOE (Department of Energy) has a goal of supporting large real-time computations, such as data processing due to an astronomical event, a high-energy particle accelerator, or responding to natural disasters.
This requires preemptible jobs using system-level checkpointing on short notice (even minutes).
An example of large real-time computations with a particularly extensive literature is X-ray scattering experiments using Free Electron Lasers
(XFELs)~\cite{blaschke2021real,blaschke2023lightsource,giannakou2021experiences}.
Still other examples can be found in the workshop on \emph{Interactive and Urgent HPC} (\url{https://www.urgenthpc.com/}).  Library-based checkpointing using the main-loop execution model may not be able to reach the next globally synchronized iteration within the short notice.  Quoting from~\cite[Section3.6]{bard2022lbn} (Real-time scheduling):
\begin{quotation}
\noindent
``In the next five years, we expect an increasing amount of our workload to come from
short-notice, rapid-turnaround compute jobs from experiment facilities. Our system of
reservations plus preemptible jobs is working well at relatively small scales - it is unclear
whether this will scale to jobs that require a significant fraction of the machine. Work will be
needed to appropriately incentivise a preemptible workload to fill the gaps in reservations. This
will include increased use of (and support for) checkpointable applications.
$\ldots$
A large pool of
general workload that is preemptible (e.g., via user or perhaps system checkpointing) may be
an option, but significant work is still needed to identify a technical solution.''
\end{quotation}

\subsection{MANA's New Implementation-Oblivious Virtual Ids: \\
            \hspace*{0.4cm}Supporting diverse MPI implementations}

MANA's new production platform is now capable of supporting transparent checkpointing for any standards-compliant implementation of MPI,
while maintaining low runtime overhead.  Low overhead is essential for production MPI jobs.  This work is based on a flexible design for virtual ids that is no longer targeted primarily toward HPE~Cray~MPI (based on MPICH~\cite{gropp1996user}).  This is demonstrated on the highly diverse MPI implementations by MPICH~\cite{gropp1996user}, Open~MPI~\cite{graham2006open}, and ExaMPI~\cite{skjellum2020exampi}.  For details, see Section~\ref{sec:background:splitProcess} for MANA's need for virtual ids, and Section~\ref{sec:virtualIds} for a novel virtual-id design to implement a platform supporting arbitrary standards-compliant implementations of MPI.

As transparent checkpointing for MPI reaches ever wider use, it is
important to accommodate multiple MPI implementations.  Developers
should not need to port MANA to work with each new major release of a
MPI implementation.  The existing MANA code is unfortunately directly
tied to implementation decisions within HPE~Cray~MPI (hereafter called Cray~MPI). This work
concentrates on the novel implementation-oblivious production platform needed to bring that vision
to fruition.

The ability of MANA developers to ``develop once, run everywhere'' impacts strongly on user support.  As an example, MVAPICH~\cite{gao2006application} and Open~MPI~\cite{hursey2007design,hursey2009interconnect} both used to include separate support for transparent checkpointing.
Both MPI implementations were later forced to drop support for transparent checkpointing, due to the large developer burden of supporting an ever-increasing variety of network interconnects. 
By adopting a philosophy of ``develop once, run everywhere'', MANA hopes to avoid that trap.

MANA~\cite{garg2019mana} demonstrated ``MPI-Agnostic Network-Agnostic''
checkpointing in an academic prototype in 2019.  Since then, the
academic prototype has been made fully robust, and is planned to
enter production use at NERSC in early Fall of 2023.
MANA's novel split-process design originally promised the
``MPI-Agnostic'' feature:  the promise to \emph{develop MANA once,
and use MANA with every MPI implementation}.
However, this feature was only tested in the original academic
prototype, which concentrated on Cray MPI.
An analysis of the original MANA internals
shows that much of the original design was hardwired for Cray MPI.

This work describes a new platform for MANA that is
MPI ``implementation-oblivious'', to distinguish from
the more limited MPI-agnostic demonstration of the original
academic prototype.  The prototype was tested
only on Cray~MPI, with the exception of one experiment checkpointing
under Cray~MPI, and then restarting under Open~MPI~\cite[Section~3.6]{garg2019mana}.
That experiment ran a simple version of GROMACS~\cite{gromacs}, which was restricted
to the MPI primitives (such as \texttt{MPI\_COMM\_World}),
and did not create any new MPI objects --- not even a new
MPI communicator.  Finally, when the academic prototype was
replaced by a production-quality version, even that
limited version of MPI agnosticism was lost.
(To understand better why the original GROMACS experiment~\cite[Section~3.6]{garg2019mana}
cannot be extended to larger MPI codes, see Section~\ref{sec:futureWork}.)

This work demonstrates a modified MANA that is truly ``MPI-Agnostic''.
It is tested against Cray~MPI, MPICH, Open~MPI, and ExaMPI.
The previous MANA work replaced the communicator, group, request, operation and datatype
integer ids of Cray MPI and MPICH with virtual ids.  The use of MANA-internal
virtual ids allows MANA to re-bind the virtual ids to new ids in the MPI library
during restart.  However, the 32-bit virtual integer ids are not compatible
with the 64-bit pointers and dynamically allocated MPI global constants used by Open~MPI and ExaMPI.

In this work, a new ``virtual-id'' subsystem was designed to eliminate the multiple
places where the original MANA design was hardwired to favor Cray~MPI.
Further, the new design is also intended to support subsets of the
full MPI implementation (e.g., ExaMPI).  For this purpose, a modified version
of MANA was developed to (i)~use the new virtual-id subsystem; and
(ii)~remove MANA's reliance on some MPI calls outside of a core subset.
This work also identifies the core subset of MPI required for MANA support.

\medskip\noindent
\subsection{Points of Novelty}
\hfill\newline
The novelty of this work is as follows.
\begin{enumerate}
    \item The Open MPI and ExaMPI implementations do not currently support transparent checkpointing.  A single, implementation-oblivious
           codebase now supports transparent checkpointing for Open~MPI and ExaMPI, as well as continuing to support the MPICH family.
    \item  This design makes MANA more flexible in supporting future MPI implementations.  A virtual id is now a pointer to a MANA-internal struct that can flexibly point to MPI physical ids based on int, pointer, pointer to struct, or other datatypes for MPI ids.  The MANA virtual id occupies the first 32 bits of any MPI id type declared in the \texttt{mpi.h} include file of a particular MPI implementation.
    \item A single virtual id that encodes any of the five MPI id types:  communicator, group, request, operation, and datatype.
    \item The MANA virtual id stores additional MANA-internal information to adapt to future evolutions of the MANA architecture (e.g., the choice between record-replay of MPI objects during restart; or use of MPI functions to serialize a representation of the MPI object; or hybrids of the two strategies).
\end{enumerate}

\medskip\noindent
\subsection{Organization of This Work}
\hfill\newline
This work is organized into the following sections.
Section~\ref{sec:background} briefly describes the underlying split-process design of the original MANA, and then provides further details of how wrapper functions are used to translate MPI calls.
Section~\ref{sec:designSpace} reviews the range of design choices made by different MPI implementations.
Section~\ref{sec:virtualIds} describes the new design of virtual ids, and how it supports multiple MPI implementations.
Section~\ref{sec:mpiSubset} specifies the MPI subset support required of a particular MPI implementation, to be supported by MANA.  The chosen MPI must support both MANA and the targeted application.  (MANA itself supports most of the full range of MPI-3.0, except for MPI's one-sided communication.)
Section~\ref{sec:experiment} presents an experimental evaluation of the novel virtual-id design for MANA, while maintaining low runtime overhead.
Section~\ref{sec:relatedWork} presents related work.
Section~\ref{sec:conclusion} is the conclusion, and Section~\ref{sec:futureWork} describes future work.

\section{Background}
\label{sec:background}

%\fix{Comments: I feel like we need a bit more background info about how MANA works at high level. e.g., it is a plugin to DMTCP and uses its checkpointing package, and mention its centralized coordinator up front at least once. I guess we should not assume everyone knows that DMTCP is a coordinated checkpointing.}

%\fix{Should we just use a single {\bf address space} instead of {\bf virtual memory}? I am not sure we use virtual memory or not on Cray systems}

\subsection{Design of MANA}

MANA (MPI-Agnostic Network-Agnostic transparent checkpointing tool) is a previously developed package for checkpointing MPI applications~\cite{garg2019mana}.  A newer version, MANA-2.0~\cite{xu2021mana}, has been developed for production use in supercomputing.  MANA uses the idea of split processes.

Hence, the large family of MPI calls are passed to the actual MPI library:
while maintaining efficiency; and
while guaranteeing
that no MPI process is blocked in a call to the lower half
at the time of checkpoint.

\subsection{Split processes}
\label{sec:background:splitProcess}

The most difficult challenge of MANA is to be able to checkpoint
an MPI application, while not having to checkpoint the network
libraries or operating system kernel modules for RDMA-based
shared memory across computer nodes.  The solution is a
split-process strategy in which the MPI-based code is split
into two parts:
\pagebreak

\begin{description}
   \item[\emph{upper half:\/}] \phantom{half:\/} the MPI application itself; and
   \item[\emph{lower half:\/}]  \phantom{half:\/} the MPI library,
                        along with the network library, kernel drivers, etc.
\end{description}
This approach is formalized using a split process architecture.

\begin{figure}[!ht]
\centering
\includegraphics[width=0.7\columnwidth]{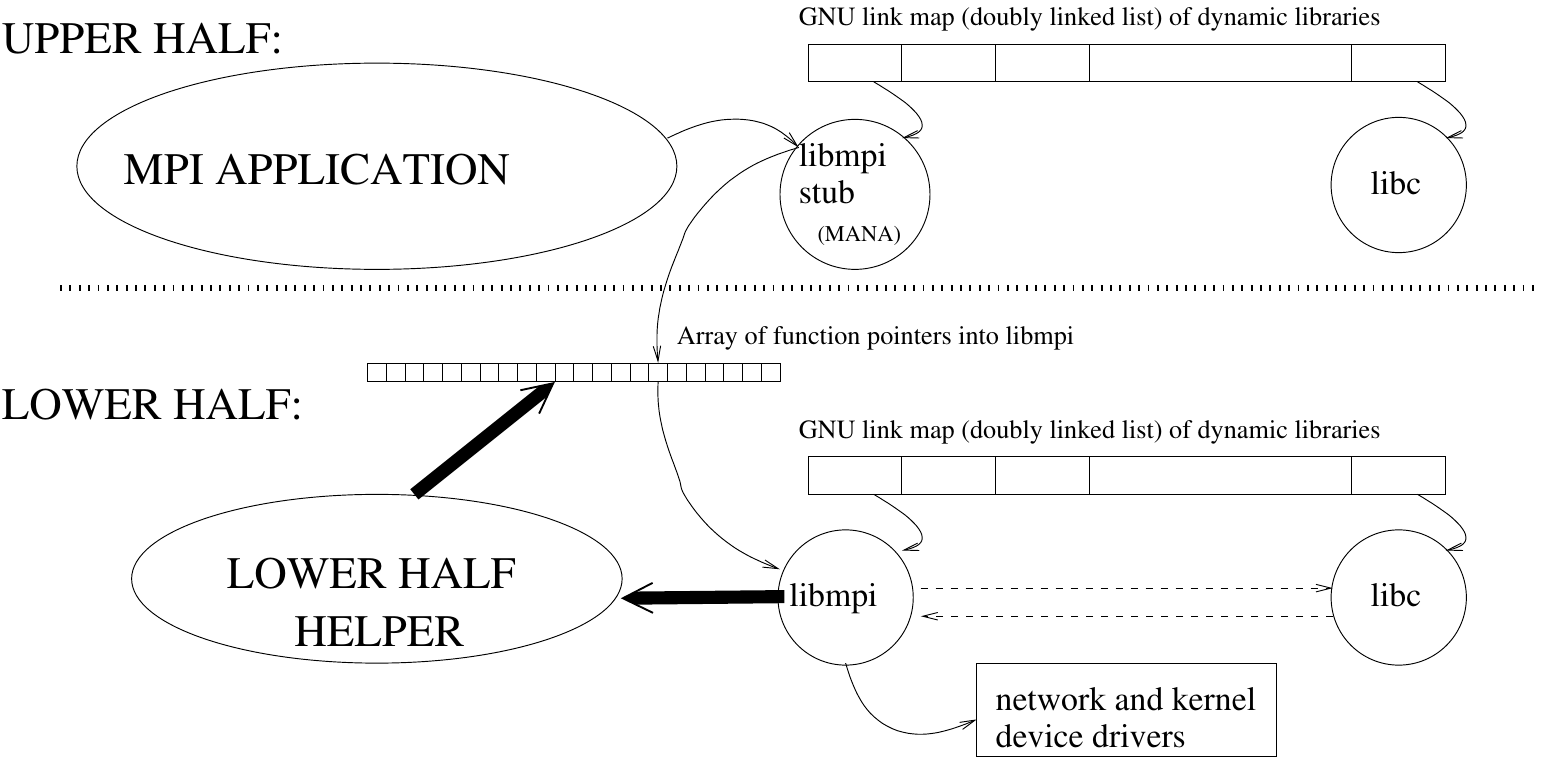}
\caption{MANA implementation:  split processes}
\label{fig:mana}
\end{figure}

The \emph{split-processes} architecture is summarized in Figure~\ref{fig:mana}.  In split processes, two programs are loaded into the address space of a single process.  One program (upper half) is the MPI application (linked to a MANA library), and the second program (lower half) is a small MPI application that includes the actual MPI library.  A MANA-internal library includes a stub function (wrapper function) for each MPI function, and each wrapper function calls one or more actual MPI functions in a small (lower-half) MANA-internal MPI program.  On checkpoint, only the memory of the upper-half MPI application is saved.  On restart, a new lower-half MPI program is launched, and that program restores the upper-half MPI application to its original location in memory.

The split process technique saves the memory of an MPI application (the upper half), while at the same time avoiding the need to save the memory of the MPI library.  This is critical, since the MPI library intensively uses hardware associated with the network and possibly a high-performance network switch.  Saving and restoring the state of the network switch, and possibly some kernel modules, is impractical, due to the difficulty of restoring the state of the associated hardware.  The term ``Network-Agnostic'' in the MANA acronym (MPI-Agnostic Network-Agnostic checkpointing) is based on using split processes to make the design of MANA independent of the particular network hardware or software drivers.

\subsection{Virtual Ids for Split Processes}
\label{sec:background-virtualIds}
A final piece of this puzzle is the problem of restoring the ids for MPI communicators, groups, and datatypes at the time of restart.  In the split-process design, the MPI application is linked to a stub library, which calls the actual MPI library function in the lower half (see Figure~\ref{fig:mana}).  Upon creation of a communicator, group, request, operation, or datatype, a corresponding global MPI id is created by the lower-half MPI library and passed to the end-user code (the upper-half MPI application).

Upon restart, before returning control to the user's MPI application, MANA makes calls to the lower-half MPI library to create new copies of each of the communicators, groups, and datatypes that were present at the time of checkpointing.
The difficulty here is that on restart, the MPI library will create new MPI ids that do not correspond to the original MPI ids at the time of checkpoint.  Hence, any MPI ids saved and later restored inside the memory of the upper-half MPI application will no longer be valid.

The solution is to maintain a MANA-internal table of virtual MPI ids that are passed to the upper half and physical MPI ids that are maintained in the lower-half MPI library.  Thus, in Figure~\ref{fig:mana}, the stub function shown in the upper half refers to the MANA-internal table and translates from virtual id to physical id when making an MPI function call; and translates from physical id to virtual id when returning from the MPI function call.

The core issue in this work is to design a more general scheme for virtual-to-real translation of ids.  The ids include MPI ids for communicator, group, request, operation, and datatype.  During runtime, each MPI application can create new instances of ids.

\section{Design choices made by different MPI implementations}
\label{sec:designSpace}

This section reviews the range of design choices made by different MPI implementations.

Communicators/datatypes in Open~MPI are 64 bits, for sake of a pointer.
The pointer directly points to an internal struct with information
about the communicator, group, request, operation or other.

The MPICH family of MPI implementations (including MPICH, MVAPICH, Intel~MPI, and Cray~MPI)
uses a special 32-bit id to support a 2-layer table
(similar to what is found in 2-layer page tables in operating systems):
The MPICH id consists of: (i)~bits representing whether it is a communicator,
group, or other; (ii)~first-level index into an internal table that then
points to a second table; and (iii)~the second-level index that accesses
the actual struct representing the communicator, group, request, operation, or other MPI objects.

ExaMPI makes an unusual design choice.
Primitive datatypes are defined in an enum class.
Other types are defined as pointers to internal structs.
The enum conflicts with the class template used by MANA.  The new
virtual-id subsystem is compliant with the ExaMPI design choices.

\section{A New Architecture for Virtual Ids of MPI Objects}
\label{sec:virtualIds}

First, we present the motivation for why a new architecture
is needed for MANA's virtual ids.  In the following
subsection, we present the new architecture that now
supports MANA's MPI-oblivious feature.

\subsection{Motivation}
\label{sec:motivation}

MANA's current design of virtual id's for MPI types like
MPI\_Comm is based on a series of C++ \texttt{std::map}. Each map represents one
MPI type.  This design has several drawbacks:
\begin{enumerate}
    \item All virtual ids are defined as \texttt{int}. It's okay when running with
MPICH. However, other MPI implementations define MPI types as pointers to
their internal structure representations. Using int as virtual id's will
conflict with user applications linked with other MPI implementations.
    \item MANA's
existing virtual ids use C++ associative arrays based on a string index.
Repeated string comparisons add some small overhead.
    \item The virtual id table only translates between real and virtual ids
without any other information. As a result, all data associated with a
virtual id have to be stored in separate maps. When accessing multiple
data related to the same virtual id,
MANA needs to look up the same virtual id multiple times.
    \item MANA needs to replay functions that create communicators, groups, operations, requests,
and datatypes on restart to update the real id's in the virtual id table.
    \item The virtual-to-real translation's performance depends on C++'s map
implementation.  \texttt{std::map} is O(logn) and \texttt{std::unordered\_map} is faster
O(1). However, the real-to-virtual translation is always O(n) because
MANA needs to iterate over all values of the map.  (Real-to-virtual translation
is used rarely --- in just one MANA wrapper function.)
\end{enumerate}

\subsection{The New Virtual Id Architecture}
\label{sec:new-vid}

To solve the problems mentioned above, a new data structure has been designed,
based on a two-level table.  Prior to this work, MANA maintained a
simple integer for the virtual id, with a separate virtual id
for each type of MPI object id.  In the academic
prototype of MANA~\cite{garg2019mana}, this was sufficient to support Cray MPI.
But as MANA grew to production quality,
independent data structures had to be added, motivated solely by support
for Cray MPI.  Hence, MANA's initial promise of being MPI-agnostic was
lost in the production development.  A new virtual id architecture
presented here now recovers that MPI-agnostic feature, and is here designated
implementation-oblivious, to recognize the new architecture.

Each virtual id in the new design is represented by a structure that
corresponds to an MPI communicator, group, request, operation, or datatype.
This structure contains additional MANA-specific information associated
with that MPI object.  The MANA-specific information can be updated
during normal execution.  It is used to correctly save the state of
MPI objects created by the lower-half MPI library.

The MANA virtual id consists of a 32-bit integer index into a
table of the structures described earlier.  The 32-bit index
internally includes a \emph{ggid} or ``global group id'' for
communicators and groups, and a related index for request,
operation, and datatype.
The MANA stub functions of the upper half pass back
to the MPI application the MPI object id (MPI\_Comm, MPI\_Group, etc.)
as declared by the include file for the specific MPI implementation.
MANA embeds its virtual id (the 32-bit integer) into the first 4~bytes of
the MPI object type declared by the MPI include file.  The integer
is an index into a table of pointers to the structures.  When the
MPI application (compiled against the existing MPI include file)
passes an MPI object as an argument, the MANA stub function recovers
the 32-bit integer and uses this to call the lower-half MPI
library with the corresponding ``physical'' MPI object id.

The new virtual-id structures are important to MANA at the time of checkpoint and restart.
At the time of checkpoint, the structures may be further updated based
on the current state of MPI objects.  The structures are then saved
as part of the checkpoint image of the upper half.  MANA does not
require a special data structure in the checkpoint image to identify
these MANA-internal structures.

At the time of restart, MANA must create MPI objects that are semantically
equivalent to the objects that existed prior to checkpoint.  The MPI objects
are re-created during restart.  After restoring the upper-half memory,
control is passed back to MANA in the upper half. For each MANA-internal
structure representing an MPI object, MANA calls the MPI library in the
lower half, in order to create a semantically equivalent MPI object.
MANA then updates the internal structures to represent the new
``physical '' object id (of types MPI\_Comm, MPI\_Group, etc.) returned
by the call to the MPI library of the lower half.

Note that a consequence of the design of MANA (regardless of virtual ids) is that MANA must be
recompiled for each MPI implementation that it supports.  This is
required because the MANA stub functions (see Figure~\ref{fig:mana})
make calls to the MPI library, and MANA itself may make internal
calls to the MPI library.  Hence, MANA must be recompiled against
the MPI include file for the targeted MPI implementation.  However,
under the new virtual-id system, none of the code of MANA needs to be 
modified when a different MPI include file (``mpi.h'') is substituted.  
Hence, MANA remains truly ``MPI implementation-oblivious''.

\subsection{MPI Global Constants as functions}

In the development of virtual ids to support MPI implementations outside
the MPICH family, it was discovered that Open~MPI implements certain
global constants, such as MPI\_COMM\_WORLD, as macros that expand
to functions.  In the MPICH family, MPI\_COMM\_WORLD and other
MPI constants expand to
unique integers that are the same in the upper and lower half, and
the same before checkpoint and after restart.

In Open~MPI, it expands to a function call that returns a pointer,
which may vary when it occurs in the upper-half MPI application
(linked dynamically against MPI),
or in the lower-half trivial application (linked statically against
MPI).  (See Figure~\ref{fig:mana} to review the MANA architecture.)
Further, the value of MPI\_COMM\_WORLD for Open~MPI can vary
between its use prior to checkpoint and its use after restart.

ExaMPI adds a small additional requirement to MANA.  In Open~MPI,
global MPI constants are determined at library startup.  But ExaMPI
defines the global constants using smart, shared pointers with
reinterpret casts.  This is done so that, for example, MPI\_INT8\_T
and MPI\_CHAR can share a pointer.  But a consequence of this decision
is that the address of a constant is known relatively late at runtime,
on a lazy basis.  MANA now accounts for this ``lazy'' design, when
translating MPI global constants
in the upper half to the correct lower-half form.

The situation is analogous to the case of MPICH's Fortran named constants
(e.g., MPI\_STATUS\_IGNORE, MPI\_MPI\_IN\_PLACE),
which can vary from one session to the next~\cite[Section~3.3]{zhang2014implementing}.
The solution in the case of MANA is to re-define the MPI\_COMM\_WORLD and other macros
as a pointer to a lower-half array that contains the results of
calling the corresponding functions.

\section{MPI Subset Requirements for ``MPI-Agnostic'' MANA Support}
\label{sec:mpiSubset}

MANA uses MPI functions internally to operate the network and to gather and sync MPI's runtime status among ranks. MANA is designed to be MPI implementation agnostic and network agnostic. Therefore, MANA cannot use lower-level network libraries for operations like draining messages in the network.

MANA requires three categories of MPI functions from MPI implementation to work properly:
\begin{enumerate}
    \item Functions that send, detect, and receive messages in the network. At checkpoint time, some point-to-point messages may still be pending in the network. MANA requires MPI\_Iprobe to detect pending messages in the network, MPI\_Recv, and MPI\_Test to complete pending point-to-point communications.
    \item Functions that decode MPI objects that need to be reconstructed at restart time. MPI objects like communicator, operation, and datatype need to be reconstructed at restart time. Therefore, MANA needs the functions that decode such objects to obtain essential information for the reconstruction. Currently, MANA requires: MPI\_Comm\_group, MPI\_Group\_translate\_ranks, MPI\_Type\_get\_envelope, and MPI\_Type\_get\_contents.
    \item A small set of MPI communication functions used by MANA to share messages among processes, which includes: MPI\_Send, MPI\_Recv, and MPI\_Alltoall.
\end{enumerate}

\section{Experimental Evaluation}
\label{sec:experiment}

%% Some smaller applications from SPEC MPI2007 are here:
%%   https://www.spec.org/mpi2007/Docs/faq.html#Measures
%%   Are they useful?
%% In the future, it might be fun to download the
%% SPEC MPI2007 for free (non-profit), and then to run under
%% MANA and report some SPEC times.
%%%% https://www.spec.org/mpi2007/results/

The experiments for MPICH, Open MPI and ExaMPI were run on a local HPC cluster called ``Discovery'',
at Northeastern University, using Linux kernel~3.10
and TCP for all MPI.
\todo{(InfiniBand is available, but it would take longer to fix that for virtId.)}
The cluster was configured with CentOS~7.9 and Slurm~21.08.8-2.
Jobs were run on Cascade Lake processors:  Intel Xeon Platinum 8276 rated at 2.20GHz.
Each node had 2 sockets, with a total of 56~cores per node (no hyper-threading).
The MPI implementations were MPICH-3.3.2 (provided on the cluster);
Open~MPI~4.1.5 (built locally, since both dynamically and statically linked MPI are required
to build MANA); and ExaMPI (git developer branch for August,~2023, requiring C++-20).
The ``mpicc'' commands were based on gcc-10.1.
Due to the experimental nature of ExaMPI, it was tested with a subset of applications known to be compatible. In all instances, MANA was built with gcc optimizations enabled. Timings were obtained through the use of SBATCH scripts and the Linux \texttt{date} utility -- internal application timings were NOT used.

\todo{We're seeing large standard deviation for Cray~MPI and for ExaMPI.  We need to rerun some of those for longer periods of time, to see if we can any remove high variance associated with startup effects.}

Section~\ref{sec:craympi} contains runs on the Perlmutter supercomputer,
using Cray~MPI and dual-socket AMD EPYC 7763 CPUs.  Cray mpicc is based on gcc-11.2.
The Linux operating system is SUSE Linux Enterprise Server 15 SP4 (Release~15.4),
with Linux kernel~5.14.  The Perlmutter supercomputer is the \#8 supercomputer,
as of the TOP500 list of June, 2023~\cite{top500jun2021}.

%, and running over Broadwell processors (28 cores split among 2~sockets.
%for Intel Xeon CPU E5-2680 at 2.40GHz), instead of Cascade Lake.
%The Broadwell processors were chosen for testing over TCP, since ExaMPI's InfiniBand support
%is still experimental.

\todo{We still should experiment with how to use mana\_launch with 2 nodes, properly.  But maybe
	we'll do this later, if the paper is accepted.}

Note that due to the much older Linux-3.10 kernel (introduced in~2013) at the local site,
the kernel does not support the FSGSBASE kernel feature (introduced in~2020).
Hence, the MANA split processes required using a call to the kernel (``prctl(ARCH\_SET\_FS, ...)'')
to switch to a new ``fs'' register during any MPI function calls.  The FSGSBASE feature
allows recent kernels to support switching ``fs'' registers in user space with a
single assembly instruction.  The penalty of using the older ``prctl'' approach has been found to range anywhere
from 3\% to 30\% or higher, depending on the frequency of MPI calls (calls to the lower-half program).

%The selected applications include CoMD~\cite{papa2001constrained,mohd2013co}, HPCG~\cite{dongarra2016new}, LAMMPS~\cite{thompson2022lammps}, Lulesh-2.0~\cite{karlin2013lulesh},
%and SW4~\cite{petersson2015wave}.
%For a very long list of MPI-based real-world applications, see Laguna \hbox{et al.}~\cite{laguna2019large}.

Laguna \hbox{et al.}~\cite{laguna2019large} present a long list of MPI-based real-world applications. We selected five real-world applications with diverse MPI features, including ExaMPI compatibility. The selected applications include CoMD~\cite{papa2001constrained,mohd2013co}, HPCG~\cite{dongarra2016new}, LAMMPS~\cite{thompson2022lammps}, Lulesh-2.0~\cite{karlin2013lulesh}, and SW4~\cite{petersson2015wave}.

In the rest of this section, subsection~\ref{sec:openmpi} shows the use of MANA+virtId (the new virtual id design) on Open~MPI.
Subsection~\ref{sec:exampi} shows the use of MANA+virtId on ExaMPI.
Subsection~\ref{sec:craympi} shows the use of MANA+virtId on Cray~MPI on the Perlmutter supercomputer. We report runtime overhead for these cases.  Subsection~\ref{sec:ckpt-time} shows trends for checkpoint time versus checkpoint image size.

Table~\ref{tbl:inputsDiscovery} shows the inputs used for each of the applications of Figure~\ref{fig:runtimes-ompi}.

\begin{table}[ht]
	\centering
	\begin{tabular}{ |c|c|l|}
		\hline
		\textbf{App.}  & \textbf{Ranks} & \textbf{Input}  \\
		\hline
		CoMD  & 27 = 3$^3$ & \texttt{-N 10000} \\ \hline
		HPCG  & 56 & \texttt{--nx=104 --ny=104 --nz=104 --it=50 }\\ \hline
		% HPCG & --nx=104 --ny=104 --nz=104 --it=50 (104 grid points in x,y,z and 50 iterations) \\ \hline
		LAMMPS & 56  & \texttt{-in bench/in.lj (run=50000)} \\ \hline
		LULESH & 27 = 3$^3$ & \texttt{-p -i 100 -s 100} \\ \hline
		SW4 & 56 & \texttt{tests/curvimr/energy-1.in} \\ \hline
	\end{tabular}
	\smallskip
	\caption{\label{tbl:inputsDiscovery} Input for each application on a single node}
\end{table}

\subsection{Analysis of Open MPI versus MPICH}
\label{sec:openmpi}

\begin{figure*}[ht]
	\centering
	\includegraphics[width=0.9\textwidth]{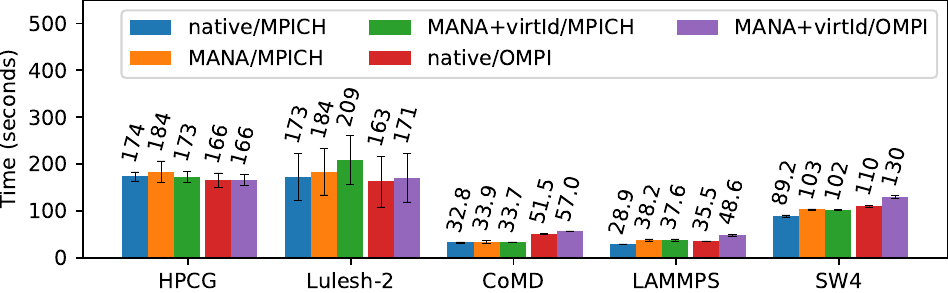}
	\caption{\label{fig:runtimes-ompi} Application runtimes of
		MPI for MPICH versus Open MPI.	A median of ten trials for each
		application was taken, and the standard deviation is shown.  The five
		cases in the legend are:  native, MANA, and MANA with virtual ids
		(virtId) for MPICH, and then native and MANA with virtual ids,
		for Open MPI.}
\end{figure*}

Figure~\ref{fig:runtimes-ompi} shows runtimes for each application in five cases:
\begin{enumerate}
	\item native/MPICH (no MANA)
	\item MANA/MPICH (the previous production version of MANA, using MPICH)
	\item MANA/MPICH with the new virtual-ids feature, described in~\ref{sec:virtualIds}
	\item native/Open-MPI (no MANA)
	\item MANA/Open-MPI with the new virtual-ids feature, described in~\ref{sec:virtualIds}
\end{enumerate}

The MPICH implementation of MPI was highlighted since
HPE Cray MPI is part of the MPICH family. HPE Cray MPI
and MPICH share much of their code. This makes possible
a rough comparison of trends, by using MPICH as the ``standard'' for comparison on the local site, and HPE Cray MPI as
the standard for production jobs on Perlmutter.
\todo{ We could run VASP for Cray~MPI on Perlmutter.  But with CoMD and LAMMPS, that's
	probably enough.}

\begin{comment}
	As seen in Figure~\ref{fig:runtimes-ompi}, HPCG and Lulesh-2 incur approximately a 5\% runtime overhead
	when using MANA.  This is assumed to be due to the FSGSBASE issue on a 10-year-old Linux kernel. We also observe that HPCG and Lulesh-2 have substantially more runtime variance than the other applications. This is speculated to be due to their large I/O writes incorporating variations in filesystem performance at the local site into the timings.
\end{comment}

As seen in Figure~\ref{fig:runtimes-ompi}, HPCG and Lulesh-2 have substantially more timing variation than the other applications even in native execution, which appeared to fall into clusters. 
\todo{I'd really love to do a hopkins test or other clustering metric to prove my point about clustering here. But maybe it's out-of-scope.}
As such, reasoning about the possible runtime overhead is more difficult -- it is unlikely that MANA+virtId can improve HPCG performance while running with MPICH beyond the native execution. This work still illustrates the new capability to run these applications under non-MPICH MPI. Runtime overhead analysis later in this section is restricted to CoMD, LAMMPS, and SW4, although there exists some evidence to believe that the true runtime overhead with HPCG and Lulesh-2 is low (Section~\ref{sec:contextswitch}).

In the cases of Lulesh-2 and SW4, a particular observation is pertinent.  These programs are
not built in the default manner (using MPI plus OpenMP). When running \emph{natively} under MPICH, Lulesh-2 consumed 50\% more time
than native Open~MPI.  This was tracked down to the MPICH on Slurm~21.08.8-2 and CentOS~7.9 thrashing with too many OpenMP threads when run with Lulesh-2.  As a workaround, Lulesh-2 is built without OpenMP support. All Lulesh-2 tests are reported in this mode. Similar thrashing behavior was observed with SW4, and the same workaround was applied.

We observe that while running with MPICH, the new virtId feature can improve performance by up to 1.6\% (in the case of LAMMPS). We speculate that this is the result of an improved differentiation and access mechanism for virtual ids, which embeds binary tags in the 32-bit virtual ids to determine virtual id type inside a single map, instead of maintaining separate singleton maps for each type chosen between via macro-encoded string comparison. Especially when very many MPI calls are made, the time needed to look up a virtual id can become a significant factor.

We observe that while runtime overhead differs for each application (see Section\ref{sec:contextswitch}), runtime overhead between OpenMPI and MPICH in each application is comparable. However, the overhead under OpenMPI tends to be greater than under MPICH. For instance, LAMMPS has a 37\% runtime overhead under OpenMPI, and 32\% under MPICH. SW4 has a 18\% runtime overhead under OpenMPI, and 15\% under MPICH. This may be the result of a difference in the speed of network calls causing more context switches to occur, e.g., while MANA internally calls \texttt{MPI\_Test} while wrapping non-blocking communication.

\subsection{Analysis of ExaMPI versus MPICH}
\label{sec:exampi}

\begin{figure}[ht]
	\centering
	\includegraphics[width=0.5\columnwidth]{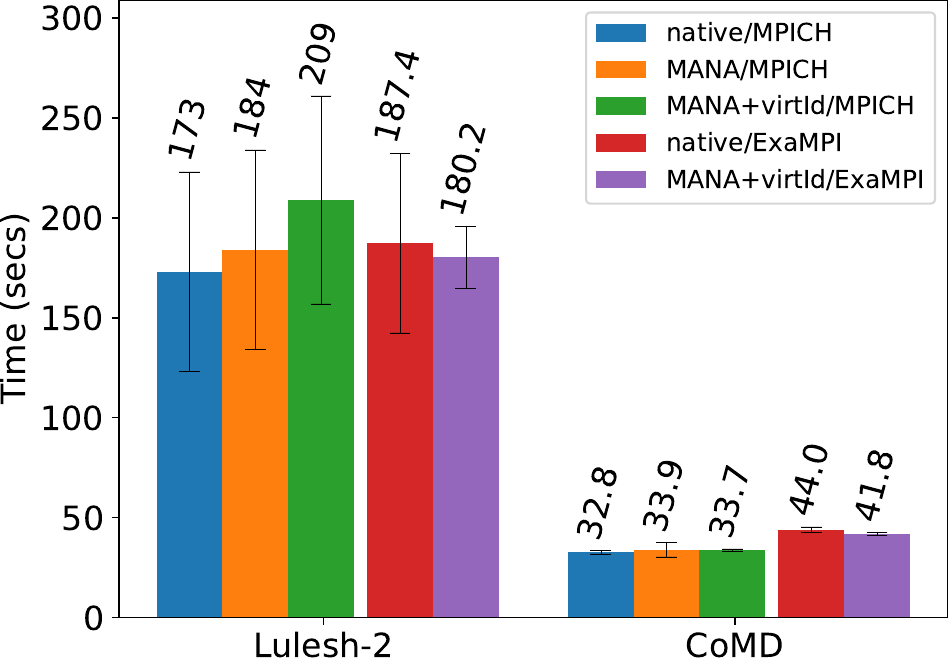}
	\caption{\label{fig:runtimes-exampi} Runtimes for ExaMPI on Discovery.
		A median of ten trials for each
		application was taken, and the standard deviation is shown.  The five
		cases of the legend are similar to that indicated
		in Figure~\ref{fig:runtimes-ompi}.}
\end{figure}

Figure~\ref{fig:runtimes-exampi} demonstrates the ability of MANA to run with ExaMPI by using the new virtId feature.  Recall that ExaMPI is intended as an experimental implementation that allows developers to \emph{easily} experiment with new algorithms.  Hence, rather than carry out research on implementing multiple, experimental new algorithms inside a production MPI, one can do the same experimentation more easily inside the smaller, C++-based ExaMPI codebase, and later port the best of the algorithms to a production implementation.

We build Lulesh-2 without OpenMP in order to be consistent with the methodology used for MPICH and OpenMPI.  Recall that native MPICH had unaccountably high overhead when built with OpenMP in the experiments of Section~\ref{sec:openmpi}.

These experiments show that MANA+virtId under ExaMPI can improve the runtime of CoMD by approximately 5\% as compared to its native runtime under ExaMPI.  It is speculated that the new virtId design may improve cache performance by producing greater code locality, or by caching some information that is otherwise re-computed in ExaMPI.  We are continuing to investigate the exact cause of the improved performance with MANA+virtId.

\todo{We can do other ExaMPI-friendly applications later, including using the missing two MPI functions.  We don't have to mention the formerly missing ExaMPI functions in this paper.  We can do the extra applications for the camera-ready version.}

\todo{Mention progress thread of ExaMPI, overlap comm. and computation.}

\subsection{Analysis of context switches per application}
\label{sec:contextswitch}

We observe that the runtime overhead measured at the local site for the five applications varies per application. There exists evidence to support the view that the body of this overhead variation is accounted for by the lack of the userspace FSGSBASE feature at the local site. Recall that this feature is a factor whenever MANA switches into the lower half (and back out of it) per the split-process architecture for MPI API wrappers.

We examined the amount of context switches performed by each application in a batch run of the graphed class, i.e., the same input and the same quantity of ranks as shown in Figure~\ref{fig:runtimes-ompi}. Although the precise number of context switches can vary (as would be expected under network variability), the quantity of switches differs by as much as an order of magnitude between applications. In one experiment, CoMD made 3.7M CS/s under 27 ranks. Under 56 ranks, LAMMPS made the most switches: 22.9M CS/s. SW4 made 12.5M CS/s. This observation agrees with the relative differences in runtime overhead observed when running these applications under MANA. Recall that, per the split-process architecture, an application making more MPI calls results in MANA performing more context switches. Therefore, with more MPI calls, the lack of userspace FSGSBASE at the local site becomes a more significant factor.

HPCG and Lulesh were also tested for context switch rate: 4.7M CS/s and 1.3M CS/s respectively. This observation implies that the true runtime overhead for these applications is comparable to that of CoMD, or yet lower.

\subsection{Analysis of FSGSBASE runtime overhead for Cray~MPI on Perlmutter}
\label{sec:craympi}

\begin{figure}[ht]
	\centering
	\includegraphics[width=0.5\columnwidth]{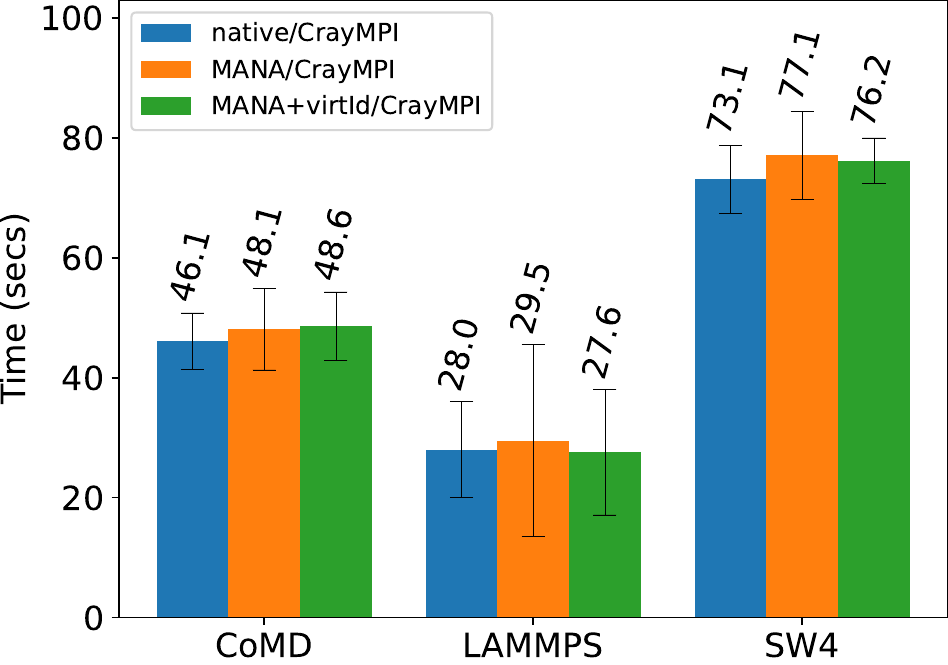}
	\caption{\label{fig:runtimes-craympi} Runtimes for Cray~MPI on Perlmutter. 
		A median of twenty-five runs for each timing was used.
		Its purpose is to show that on a production system supporting the FSGSBASE feature,
		both MANA and MANA+virtId do indeed perform comparably to the
		native execution.  Error bars represent the standard deviation among 25~trials.
		\todo{Which std. dev.?  There are 2 kinds: divide by n or by (n-1).}
		\todo{I still need to do GCC11+NODEBUG on Perlmutter. -- Leonid}
	}
	
\end{figure}

\begin{table}[ht]
	\centering
	\begin{tabular}{ |c|c|l|}
		\hline
		\textbf{App.}  & \textbf{Ranks} & \textbf{Input}  \\
		\hline
		CoMD  & 64 = 4$^3$ & \texttt{-N 30000} \\ \hline
		LAMMPS & 64  & \texttt{-in bench/in.lj (run=50000)} \\ \hline
		SW4 & 64  & \texttt{tests/curvimr/energy-1.in} \\ \hline
	\end{tabular}
	\smallskip
	\caption{\label{tbl:inputsCray} Input for each application on Perlmutter}
\end{table}

Figure~\ref{fig:runtimes-craympi} shows the results of testing CoMD and LAMMPS
on Perlmutter, where the userspace FSGSBASE feature is present.  CoMD, LAMMPS and SW4 were
selected because Figure~\ref{fig:runtimes-ompi} showed those applications to consistently
scale poorly with MANA, especially with OpenMPI, at the local site.

Figure~\ref{fig:runtimes-craympi} confirms that the high runtime overhead for
CoMD, LAMMPS, and SW4 disappears when the userspace FSGSBASE feature is available.
It's likely that the phenomenon was emphasized for SW4 and LAMMPS
due to very frequent MPI calls for those particular applications.

%%% Standard deviation figures on Perlmutter, 10 trials:
%%% Native Comd: 6.0
%%% Mana_mpich comd: 8.7
%%% Mana_mpich_vids comd: 3.0
%%% Native lammps: 3.9
%%% Mana_mpich lammps: 4.8
%%% Mana_mpic_vids lammps: 4.1
%%% Native sw4: 5.1
%%% MANA_mpich sw4: 5.8
%%% MANA_mpich_vids sw4: 6.5

Recall that, in our context switch measurements, LAMMPS had the highest context switch rate out of any application. Yet, the median runtime overhead across twenty-five trials is found to be 5.4\% in the case of standard MANA, dramatically lower than the 32.2\% observed on the local site. SW4 is measured to have an overhead of 5.5\% with standard MANA, but this is improved to 4.2\% with virtId. Improvement in median runtime was not observed under virtId with CoMD, as it was at the local site.  These results show that virtId can potentially improve performance even with userspace FSGSBASE available.

\todo{MANA has an application in the util directory that will log
	the MPI calls being made.  Using this in a future paper may provide
	some insight into why lulesh-2 improves under MANA+virtId.}

\begin{comment}
	The experiments are:
	\begin{enumerate}
		\item Perlmutter, vASP5/PdO4/Cray-MPI with 4 or 8 nodes, and as many ranks as reasonable; We want 3 cases: native, MANA, and MANA+vid  (ckpt/restart/image size are oprtional)
		\item Perlmutter:  Let's choose one of Discovery applications, and run it on Perlmutter with the same number of nodes and ranks as on Discovery, to show comparability  Again for the 3 cases, as above.
		\item 5 cases per application:  native/MPICH, MANA/noVid/MPICH, MANA/vid/MPI-X  (for MPI-X bound to each of the three MPIs)
		\item For each application, ckpt time, restart time, total size of ckpt images
		\item Ideally with 2 nodes for --constraint=zen2 --exclusive (with IB and 64 cores per node); debug/express/short partitions available.
		\item If we have to compromise to get done in time, we can go to zen2/32-cores-per-node; If that doesn't work, we can drop down to Haswell with max \# cores per rank; If that doesn't work, we lower the number of ranks on Haswell.
	\end{enumerate}
\end{comment}

\todo{The rest of this section is my notes on what's available for real-world applications.
	For the analysis, we'll want to say what types of MPI ids are used.  It's in a LaTeX comment.  ---~Gene}

\subsection{Analysis of checkpoint times per application}
\label{sec:ckpt-time}

On Discovery, the filesystem is based on NFSv3.  Table~\ref{tbl:ckptTime} shows that checkpoint image sizes and checkpoint times follow similar trends.  While it is dangerous to draw strong conclusions from the low-performance NSFv3 filesystem, it is clear that checkpoint times will continue to be modest on the high-performance filesystems of larger HPC sites.

\begin{table}[ht]
	\centering
	\begin{tabular}{|r|r|r|r|}
		\hline
		\textbf{Application}  & \textbf{Ckpt size/rank} & \textbf{Ckpt time} & \textbf{MB/s/rank} \\
		\hline
		CoMD  & 32MB & 8.9 & 3.6 \\
		LAMMPS & 42MB  & 12.8 & 3.3 \\
		SW4 & 49MB & 12.3 & 4.0 \\
		Lulesh-2 & 207MB & 16.3 & 12.7 \\     
		HPCG  & 934MB & 72.9 & 12.8 \\
		\hline
	\end{tabular}
	\smallskip
	\caption{\label{tbl:ckptTime} Checkpoint times on Discovery}
\end{table}

\section{Related Work}
\label{sec:relatedWork}

MANA~\cite{xu2021mana} appears to be the only current production-ready package supporting transparent checkpointing of MPI.

The history of checkpointing of MPI is one that moved toward greater ease of use.  Some milestones along the way include:
\begin{itemize}
    \item application-level library-based checkpointing~\cite{bronevetsky2003automated,schulz2004implementation} (2003, 2004);
   \item  MPICH-V~\cite{bouteiller2006mpich} (2006), based on a software framework at the lowest level of the MPICH~\cite{gropp1996user} software stack;
    \item BLCR~\cite{hargrove2006blcr} and DMTCP~\cite{ansel2009dmtcp} (2006, 2009) as lower-level tools for transparently saving and restoring process images;
    \item the MVAPICH2 checkpoint-restart service~\cite{gao2006application}) (2006), based on BLCR;
    \item the Open~MPI checkpoint-restart service~\cite{hursey2007design,hursey2009interconnect} (2007, 2009), based on BLCR;
    \item an early attempt at MPI-agnostic checkpointing over InfiniBand~\cite{cao2014transparent} (2014), based on DMTCP
      (including the first petascale transparent checkpoints: 16,368 processes
	for NAMD and 32,368 processes for HPCG~\cite{cao2016system} (2016));
    \item VeloC~\cite{nicolae2019veloc} (2019) with its support for asynchronous, adaptive I/O in library-based checkpointing, as well as the earlier library-based tools: SCR~\cite{moody2010design} (2010), FTI~\cite{bautista2011fti} (2011), ULFM~\cite{bland2013post,losada2020fault} (2014), and Reinit~\cite{laguna2016evaluating} (2016, a simpler interface inspired by ULFM); and
    \item MANA~\cite{garg2019mana} (2019) for network-agnostic, transparent checkpointing (but primarily focused on Cray~MPI and the MPICH family of MPI implementations);
\end{itemize}

In the quest for a production-ready checkpointing package for MPI, VeloC~\cite{nicolae2019veloc} and MANA~\cite{xu2021mana} are the two packages in greatest recent use (although SCR and ULFM/Reinit continue to see active use).  The MVAPICH and Open~MPI packages that were dependent on BLCR were sometimes used after they were developed.  But BLCR does not support SysV shared memory, and so the MVAPICH and Open~MPI checkpointing packages are no longer supported.

As noted in the introduction, the original and later work on MANA~\cite{garg2019mana,chouhan2021improving,xu2021mana} concentrated mainly on Cray~MPI, on the Cori supercomputer at NERSC~\cite{NERSC}.  An exception to the use of Cray~MPI was an example checkpointing under Cray~MPI on Cori and restarting under Open~MPI at a university cluster~\cite[Section~3.6]{garg2019mana}.  That example was performed solely on GROMACS~\cite{gromacs}, and did not create any additional MPI ids beyond those that are built into MPI (such as MPI\_COMM\_WORLD).

Cray~MPI~\cite{cray2914understanding,graham2007comparison} and MVAPICH~\cite{panda2013mvapich,panda2021mvapich} derive from the MPICH~\cite{gropp1996user} family of MPI implementations.  Open~MPI~\cite{graham2006open} is an independently developed full production version of MPI.  ExaMPI~\cite{skjellum2020exampi} is an experimental implementation of MPI designed to quickly test new algorithms, implementations, and features for MPI.

\section{Conclusion}
\label{sec:conclusion}

An implementation-oblivious feature was added to the production-quality MANA package for transparent checkpointing of MPI.  This feature enables a philosophy of ``developing once, and running everywhere'': i.e., re-compiling MANA against the MPI include file of each new MPI implementation.  This was demonstrated for three widely diverse MPI implementations:  MPICH, Open~MPI, and the experimental ExaMPI implementation.  This was demonstrated with five real-world applications.  A runtime overhead of about 5\% or less was demonstrated --- except where
Linux lacked the user-space FSGSBASE feature and frequent MPI calls were made.  A subset of the MPI standard was identified that is sufficient to support MANA on each of the three MPI implementations.

\section{Future Work}
\label{sec:futureWork}

Note that the MANA-internal
structure can store a copy of the MPI object id directly in the
MANA-internal structure.  The MPI object is declared in
an MPI-specific include file ( i.e., \texttt{mpi.h}), which is different for
each MPI implementation.
Currently, the MANA structure saves the ``physical'' MPI object id
directly on the MPI types found in the MPI include file, and MANA
is re-compiled for the MPI include file of each separate MPI implementation.
In future work, it may be possible to define a MANA version of the
MPI include file, in which MANA's MPI types are defined to have
sufficiently large size to contain any implementation-specific information.

This would recover interoperability between MPI implementations
even at the level of checkpointing.  One could checkpoint
arbitrary MPI applications under one MPI implementation and restart
under a different MPI implementation.  This was previously
accomplished only for an MPI application (a version of GROMACS)
that used only the MPI primitives, but did not create
MPI user-defined objects.

On a more modest level, we currently use an eager policy for computing
the ggid (see Section~\ref{sec:new-vid} for new communicators. 
Because some codes repeatedly create
and free communicators in a loop, we are considering the use of
a lazy or hybrid policy.

\section*{Acknowledgments}

This  work  used  the  resources  of  the National  Energy  Scientific Computing Center (NERSC) at the Lawrence Berkeley National Laboratory.  The work of the first author was partially supported by NSF Grant OAC-1740218.
We are grateful for the collaboration of Zhengji Zhao, Rebecca Hartman and William Arndt of NERSC, along with the collaboration  of MemVerge, Inc., for earlier bringing MANA to the production quality upon which the current work is based.
We would especially like to highlight the generous donations of time by Zhengji Zhao
that helped bring MANA into its production-ready phase at NERSC.  We would also like to thank Kapil Arya for suggestions on who to work around a misfeature in MANA when taking timings for checkpoint and restart.

\bibliographystyle{alpha}
\bibliography{supercheck23}

\end{document}